\documentstyle[12pt]{article}
\begin{document}
\newfont{\elevenmib}{cmmib10 scaled\magstep1}%
\newfont{\cmssbx}{cmssbx10 scaled\magstep3}
\newcommand{\preprint}{
            \begin{flushleft}
            \elevenmib Ochanomizu\, University\\
            \elevenmib Yukawa\, Institute\, Kyoto
            \end{flushleft}\vspace{-1.3cm}
            \begin{flushright}\normalsize  \sf
            YITP-98-5\\ {\tt cond-mat./9801193} \\ January 1998
            \end{flushright}}
\newcommand{\Title}[1]{{\baselineskip=26pt \begin{center}
            \Large   \bf #1 \\ \ \\ \end{center}}}
\newcommand{\Author}{\begin{center}\large \bf
            Ruihong Yue$^a$\footnote[1]{e-mail:\ {\tt yue@phys.ocha.ac.jp}%
            }\ \
            and Ryu Sasaki$^b$\end{center}}
\newcommand{\Address}{\begin{center} \it
            $^a$ Department of Physics, Ochanomizu University,
            Tokyo 112, Japan \\
            $^b$Yukawa Institute for Theoretical Physics, Kyoto
            University,\\ Kyoto 606-01, Japan \end{center}}
\newcommand{\Accepted}[1]{\begin{center}{\large \sf #1}\\
            \vspace{1mm}{\small \sf Accepted for Publication}
            \end{center}}
\baselineskip=20pt

\preprint
\thispagestyle{empty}
\bigskip
\bigskip
\bigskip
\Title{ Lax pair for $SU(n)$ Hubbard model}
\Author

\Address
\vspace{2cm}

\begin{abstract}%
\noindent
For one dimensional $SU(n)$ Hubbard model, a pair of Lax operators are
derived, which give a set of fundamental equations for the quantum inverse
scattering method under both periodic and open boundary conditions. This
provides another proof of the integrability of the model under
periodic boundary condition.
\\ \\

\end{abstract}

\newpage

Integrable strongly correlated electron systems have been an important
research subject in condensed matter physics and mathematics. One of the
significant models is the 1-D Hubbard. The exact solution was given by
Lieb and Wu \cite{LiebWu}. However, the integrability was shown twenty years
later by Shastry, Olmedilla and Wadati \cite{Shastry,OlmedillaWadati}.
The integrability and the exact solution of the system under the open
boundary condition were discussed by several authors
\cite{DeguchiYue,ShiroishiWadati}. The Lax pair was first given by Wadati,
Olmedilla and Akutsu \cite{WadatiOlmedillaAkutsu}.
Recently, Maassarani and Mathieu have constructed the hamiltonian
$SU(n)$
 XX model and proved its integrability \cite{MaassaraniMathieu}.
Considering two coupled
$SU(n)$ XX models, Maassarani succeeded in generalizing Shastry's method
to $SU(n)$ Hubbard model \cite{Maassarani1}.  Further, he solved the
Yang-Baxter equation to prove the integrability of one dimensional
$SU(n)$ Hubbard model \cite{Maassarani2}. ( It is also proved by Martins
for $n=3,4$ \cite{Martins}.)
In this paper, we apply the quantum inverse scattering
method to 1-dimensional $SU(n)$ Hubbard model and derive the explicit
form of the Lax pair, which gives another proof of the integrability.
It is worthy to note that straightforward application of the
reflection matrix method to study the problem of the integrable
open boundary in the present $SU(n)$ Hubbard model
would encounter difficulties due to the fact that the $R$ matrix  in
reference \cite{Maassarani2} does not satisfy the crossing symmetry
condition and the lack of invertibility of a certain matrix.

 The Lax pair
formalism will give an effective method for such a system.
The 1-dimensional $SU(n)$ model in the Schr\"odinger picture is given by
\begin{equation}
{\cal H}=
    \sum_{k=1}^{N}\sum_{\alpha=1}^{n-1}\left(
        E_{\sigma,k}^{n\alpha}E_{\sigma,k+1}^{\alpha n}
     +  E_{\sigma,k}^{\alpha n}E_{\sigma,k+1}^{n\alpha }
     +  E_{\tau,k} ^{n\alpha}E_{\tau,k+1}^{\alpha n}
     +  E_{\tau,k}^{\alpha n}E_{\tau,k+1}^{n\alpha }   \right)
     +  \frac{Un^2}{4}\sum_{k=1}^{N}C_{\sigma,k}C_{\tau,k},
     \label{hamiltonian}
\end{equation}
where $E^{\alpha\beta}_{a,k}$ ($a=\sigma,\tau$) is a matrix with zeros
everywhere except for a
$1$ at the intersection of row $\alpha$ and column $\beta$:
\begin{equation}
 (E^{\alpha\beta})_{lm}=\delta^\alpha_{l}\delta^\beta_{m}.
 \label{edefs}
\end{equation}
The subscripts
$a, k$ stand for two different $E$ operators at site $k$
($k=1,\ldots,N$). The $n\times n$ diagonal matrix $C$ is
defined by $C=\sum_{\alpha<n}E^{\alpha\alpha} -E^{nn}$.
The Hamiltonian enjoys the
\( (su(n-1)\oplus u(1))_\sigma\oplus(su(n-1)\oplus u(1))_\tau\)
symmetry.
The generators are
\[
J^{\alpha\beta}_a=\sum_{k=1}^NE^{\alpha\beta}_{a,k},\quad \mbox{and}
  \quad  K_a=\sum_{k=1}^NC_{a,k},\quad \alpha,\beta=1,\ldots,n-1,\quad
a=\sigma,\ \tau.
\]

In the rest of this letter we discuss various operators
 in the Heisenberg picture. When we deal with the operators
 corresponding to
 the matrices $E$ and $C$ in the Schr\"odinger picture, they are  denoted by
adding a hat
 ( $\hat{}$ ) to the corresponding matrix:
\[
\hat{\cal Q}(t)=e^{i{\cal H}t}{\cal Q}e^{-i{\cal H}t},\qquad
{\cal Q}= E\ \ \mbox{or} \ \ C.
\]
In the following, we do not indicate the time dependence of the
operators.   Applying this method to the Hamiltonian
(\ref{hamiltonian}), we  find
\begin{eqnarray}
\frac{d\hat{E}^{nn}_{a,k}}{dt}
&=&\displaystyle i\sum_{\beta<n}\left(
      \hat{E}^{\beta n}_{a,k} \hat{E}^{n \beta }_{a,k+1}
    - \hat{E}^{n\beta }_{a,k} \hat{E}^{ \beta n}_{a,k+1}
    + \hat{E}^{\beta n}_{a,k} \hat{E}^{n \beta }_{a,k-1}
     -\hat{E}^{n\beta }_{a,k} \hat{E}^{ \beta n}_{a,k-1}\right),
    \nonumber
    \\
\frac{d\hat{E}^{n\alpha}_{a,k}}{dt}
&=&\displaystyle i\sum_{\beta<n}\left(
        \hat{E}^{\beta \alpha}_{a,k} \hat{E}^{n \beta }_{a,k+1}
    + \hat{E}^{\beta \alpha}_{a,k} \hat{E}^{n \beta }_{a,k-1}\right)
- i\left(\hat{E}^{nn }_{a,k} \hat{E}^{ n\alpha}_{a,k+1}
     +\hat{E}^{nn }_{a,k} \hat{E}^{n \alpha}_{a,k-1}\right)
 -i\frac{Un^2}{2}\hat{E}^{n\alpha}_{a,k}\hat{C}_{\bar{a},k},
 \nonumber\\
\frac{d\hat{E}^{\alpha n}_{a,k}}{dt}
&=&\displaystyle i\left(
      \hat{E}^{n n}_{a,k} \hat{E}^{ \alpha n }_{a,k+1}
      + \hat{E}^{n n}_{a,k} \hat{E}^{\alpha n }_{a,k-1}\right)
    -i\sum_{\beta<n}\left(
       \hat{E}^{\alpha\beta }_{a,k} \hat{E}^{ \beta n}_{a,k+1}
      + \hat{E}^{\alpha\beta }_{a,k} \hat{E}^{ \beta n}_{a,k-1}\right)
    +i\frac{Un^2}{2}\hat{E}^{\alpha n}_{a,k}\hat{C}_{\bar{a},k},
    \nonumber\\
\frac{d\hat{E}^{\alpha\beta}_{a,k}}{dt}
&=&\displaystyle i\left(
      \hat{E}^{n\beta }_{a,k} \hat{E}^{\alpha n }_{a,k+1}
     - \hat{E}^{\alpha n }_{a,k} \hat{E}^{n \beta }_{a,k+1}
     + \hat{E}^{n\beta }_{a,k} \hat{E}^{ \alpha n }_{a,k-1}
     - \hat{E}^{\alpha n }_{a,k} \hat{E}^{ n\beta }_{a,k-1}\right),
\label{mequation}
\end{eqnarray}
where $(a=\sigma, \tau)$ and $(\bar{a}=\tau,\sigma)$.
For an infinite system, it is not necessary to specify the boundary
condition. However, one should understand $\hat{E}_{a,0}$
is equal to  $\hat{E}_{a,N}$ under the periodic boundary condition.
In an open boundary system, $\hat{E}_{a,0}$ and $\hat{E}_{a,N+1}$ in
the r.h.s. must be regarded as vanishing.

Let us first  consider the degenerate case $U=0$ (the $SU(n)$ XX
model \cite{MaassaraniMathieu}).
In this case, the equations of motion   (\ref{mequation})
decouple into two identical sets of equations for $\sigma$ and $\tau$.
Let us introduce the $L$-operator for each of them:
\begin{equation}
L_{a,k}(\lambda)
= \cos(\lambda)\,S^{(+)}_{a,k}+\sin(\lambda)\,T^{(+)}_{a,k}+ U^{(+)}_{a,k},
\label{Ldef}
\end{equation}
in which the blocks $S^{(+)}_{a,k}$, $T^{(+)}_{a,k}$ and
$U^{(+)}_{a,k}$ are defined by
\begin{eqnarray}
S^{(+)}_{a,k}&=& (\sum_{\alpha,\beta<n}
    \hat{E}^{\alpha\beta}_{a,k}E^{\beta\alpha}_{a,au} )+
 \hat{E}^{nn}_{a,k}E^{nn}_{a,au},\nonumber\\
T^{(+)}_{a,k}&=&\sum_{\alpha<n}\left(
     \hat{E}^{nn}_{a,k}E^{\alpha\alpha}_{a,au}
    +\hat{E}^{\alpha\alpha}_{a,k} E^{nn}_{a,au}\right),
      \nonumber\\
U^{(+)}_{a,k}&=&\sum_{\alpha<n}\left(
    \hat{E}^{n\alpha}_{a,k}E^{\alpha n}_{a,au}
 +  \hat{E}^{\alpha n}_{a,k}E^{n \alpha}_{a,au}\right). \label{lmatrix}
\end{eqnarray}
They satisfy various identities inherited from the definition in
terms of $E^{\alpha\beta}$'s (\ref{edefs}):
\begin{equation}
S^{(+)}_{a,k}T^{(+)}_{a,k}
=T^{(+)}_{a,k}S^{(+)}_{a,k}=S^{(+)}_{a,k}U^{(+)}_{a,k}
 =U^{(+)}_{a,k}S^{(+)}_{a,k}=0,\quad
T^{(+)}_{a,k}U^{(+)}_{a,k}=U^{(+)}_{a,k}
T^{(+)}_{a,k}=U^{(+)}_{a,k},\ldots,.
\label{idents}
\end{equation}
We also introduce
 the $M$-operator for each species:
\begin{eqnarray}
M_{a,k}(\lambda)
&=&  \sum_{\beta<n} \left\{
         A_1\hat{E}^{\beta n}_{a,k}\hat{E}^{n\beta}_{a,k-1}
        +A_2\hat{E}^{n\beta }_{a,k}\hat{E}^{\beta n}_{a,k-1})
   \right\}E^{nn}_{a,au}
   \nonumber\\
& & +\sum_{\beta<n}B\left\{
(\hat{E}^{n\beta}_{a,k}+\hat{E}^{n\beta}_{a,k-1})E^{\beta n}_{a,au}
  +(\hat{E}^{\beta n}_{a,k}+\hat{E}^{\beta
n}_{a,k-1})E^{n\beta }_{a,au}\right\}
\nonumber\\
& &\displaystyle+ \sum_{\alpha<n}\left\{
    D_1\hat{E}^{n\alpha}_{a,k}\hat{E}^{\alpha n}_{a,k-1}
   +D_2\hat{E}^{\alpha n}_{a,k}\hat{E}^{n\alpha}_{a,k-1}
     + \sum_{\beta\neq\alpha<n}D_3(
      \hat{E}^{\beta n}_{a,k}\hat{E}^{n\beta}_{a,k-1}+
      \hat{E}^{n\beta}_{a,k}\hat{E}^{\beta n}_{a,k-1})\right\}
      E^{\alpha\alpha}_{a,au}
      \nonumber\\
& & +
\sum_{\alpha<n}\sum_{\beta\neq\alpha<n}F\left\{
   \hat{E}^{n\beta}_{a,k}\hat{E}^{\alpha n}_{a,k-1}-
   \hat{E}^{\alpha n}_{a,k}\hat{E}^{n\beta}_{a,k-1}\right\}
   E^{\beta\alpha}_{a,au},\label{mmatrix}
\end{eqnarray}
where $ A, B, D, F$ are  as yet undetermined functions of a (spectral)
parameter $\lambda$. The matrices $E_{a,au}$ are the constant matrices with
the same definition as $E_{a,k}, k=1, \cdots, N$.  The subscript $au$
stands for the auxiliary space instead of the  quantum space.
Thus $L_{a,k}$ and $M_{a,k}$ are $n\times n$ matrices in the
auxiliary space.
We want to rewrite equations (\ref{mequation}) in a matrix Lax pair form
\begin{equation}
\label{xlaxpair}
\frac{dL_{a,k}(\lambda)}{dt}=M_{a,k+1}(\lambda)L_{a,k}(\lambda)
-L_{a,k}(\lambda)M_{a,k}(\lambda).
\end{equation}
Substituting equations (\ref{Ldef}) and (\ref{mmatrix}) into the above
Lax-pair form (\ref{xlaxpair}), we find the solution
\begin{equation}
\begin{array}{c}
A_1=D_1=i+i\tan(\lambda),\quad  \quad  A_2=D_2=i-i\tan(\lambda),\\[3mm]
D_3=i,\quad\quad B=-i/\cos(\lambda), \quad\quad F=i\tan(\lambda).
\end{array}
\end{equation}
  From now on we do not denote the $\lambda$-dependence of $L_{a,k}$ and
$M_{a,k}$ for brevity.
The transfer matrix for $SU(n)$ XX model with $N$ sites can be defined
by
\[ T_{a,XX}=L_{a,N}\cdots L_{a,1},
\]
 which satisfies
\[ \frac{dT_{a,XX}}{dt}=M_{a,N+1}T_{a,XX}-T_{a,XX}M_{a,1}.
\]
  From this it is standard to show that the trace of $T_{a,XX}$ is
independent of time
under periodic boundary condition. So the $SU(n)$ XX
model is integrable.
This Lax-pair is very important for the construction of the Lax-pair of
$SU(n)$ Hubbard model. Notice that its integrability was first given by
Maassarani and Mathieu in the framework of Yang-Baxter relation
\cite{MaassaraniMathieu}.

Now, let us consider $ U\neq 0$ case. In terms of the above $L_{a,k}$ and
$M_{a,k}$, we can rewrite  equations (\ref{mequation}) as
\begin{equation}
\frac{dL_{a,j}}{dt}=M_{a,j+1}L_{a,j}-
 L_{a,j}M_{a,j}+i\frac{Un^2}{4}[L_{a,j}, C_{a,au}\hat{C}_{\bar{a},j}],
\end{equation}
in which the last term in the r.h.s. manifests the coupling between the
two species $\sigma$ and $\tau$.
Using the relation
$[L_{a,j}, C_{a,au}]=-[L_{a,j}, \hat{C}_{a,j}]$, we obtain
\begin{equation}
\label{twol}
\frac{dL_{a,j}}{dt}=M_{a,j+1}L_{a,j}-
 L_{a,j}M_{a,j}-i\frac{Un^2}{4}[L_{a,j}, \hat{C}_{a,j}\hat{C}_{\bar{a},j}].
\end{equation}
We define the following operators for the coupled system
\begin{equation}
\widetilde{L}_j=L_{\sigma,j}L_{\tau,j},\quad\quad
\widetilde{M}_{j}=M_{\sigma,j}+M_{\tau,j},
\end{equation}
then equation (\ref{twol}) can be written as
\begin{equation}
\label{onel}
\frac{d\widetilde{L}_{j}}{dt}
=\widetilde{M}_{j+1}\widetilde{L}_{j}
 - \widetilde{L}_{j}\widetilde{M}_{j}
 -i\frac{Un^2}{4}[\widetilde{L}_{j}, \hat{C}_{\sigma,j}\hat{C}_{\tau,j}].
\end{equation}
Now, we want to rewrite the last term in the above equation so that
we could obtain  the
Lax-pair form for the coupled system in a  similar form to
(\ref{xlaxpair}). Following  the  method given by Wadati et al for
$SU(2)$ Hubbard model, we introduce a ``rotation'' matrix by the
$u(1)\otimes u(1)$ charge:
\begin{equation}
I_{au}=\cosh(h/2)+\sinh(h/2)C_{\sigma,au}C_{\tau,au}
   =\exp\{{h\over2}C_{\sigma,au}C_{\tau,au} \},
   \label{iaudef}
\end{equation}
where $h$ is a free parameter to be determined later and
the ``rotated'' operators are
\begin{equation}
{\cal L}_j=I_{au}\widetilde{L}_jI_{au},\quad\quad
{\cal M}_j= I_{au}^{-1}\widetilde{M}_jI_{au}.\label{calm}
\end{equation}
By this the fundamental blocks $S^{(+)}_{a,k}$, $T^{(+)}_{a,k}$ and
$U^{(+)}_{a,k}$ are mapped to their `anti-symmetric' counterparts
$S^{(-)}_{a,k}$, $T^{(-)}_{a,k}$ and $U^{(-)}_{a,k}$:
\begin{eqnarray}
S^{(-)}_{a,k}&=&C_{a,au}S^{(+)}_{a,k}
  =S^{(+)}_{a,k}C_{a,au}=(\sum_{\alpha,\beta<n}
    \hat{E}^{\alpha\beta}_{a,k}E^{\beta\alpha}_{a,au} )-
 \hat{E}^{nn}_{a,k}E^{nn}_{a,au},\nonumber\\
T^{(-)}_{a,k}&=&C_{a,au}T^{(+)}_{a,k}=T^{(+)}_{a,k}C_{a,au}=\sum_{\alpha<n}
\left( \hat{E}^{nn}_{a,k}E^{\alpha\alpha}_{a,au}
      -\hat{E}^{\alpha\alpha}_{a,k} E^{nn}_{a,au}\right),
      \nonumber\\
U^{(-)}_{a,k}&=&C_{a,au}U^{(+)}_{a,k}
=-U^{(+)}_{a,k}C_{a,au}=\sum_{\alpha<n}\left(
    \hat{E}^{n\alpha}_{a,k}E^{\alpha n}_{a,au}
 -  \hat{E}^{\alpha n}_{a,k}E^{n \alpha}_{a,au}\right),
 \label{minblocks}
\end{eqnarray}
which also satisfy identities similar to those given in (\ref{idents}).
Then  equation (\ref{onel}) becomes
\begin{equation}
\frac{d{\cal L}_{j}}{dt}=I^2_{au}{\cal M}_{j+1}I^{-2}_{au}{\cal L}_{j}-
 {\cal L}_{j}{\cal M}_{j}
-i\frac{Un^2}{4}[{\cal L}_{j}, \hat{C}_{\sigma,j}\hat{C}_{\tau,j}].
\end{equation}
Using the definition of $I_{au}$ and ${\cal M}$, we obtain
\begin{eqnarray}
I^2_{au}{\cal M}_{j+1}I^{-2}_{au}&=&\displaystyle
{\cal M}_{j+1} +Q_{j+1}+Q_{j},\label{i2mi2}\\
Q_j&=&-\frac{2i}{\cos(\lambda)}\sinh(h)\left(U^{(-)}_{\sigma,j}C_{\tau,au}+
U^{(-)}_{\tau,j}C_{\sigma,au}\right),\label{q}
\end{eqnarray}
and
\begin{equation}
\begin{array}{rcl}
\displaystyle \frac{d{\cal L}_{j}}{dt}&=&\displaystyle
({\cal M}_{j+1}+Q_{j+1}){\cal L}_{j}- {\cal L}_{j}({\cal
M}_{j}+Q_j)
   \\[3mm]
& &\displaystyle + {\cal L}_{j}Q_j
+ Q_j{\cal L}_{j} -i\frac{Un^2}{4}[{\cal L}_{j},
\hat{C}_{\sigma,j}\hat{C}_{\tau,j}].
\end{array}
\end{equation}
Detailed calculation shows  the last line in the above equation
to be
\begin{eqnarray}
 & & Q_j{\cal L}_j+{\cal L}_jQ_j
  -i\frac{Un^2}{4}[{\cal L}_j,
  \hat{C}_{\sigma,j}\hat{C}_{\tau,j}]\nonumber\\
&=&\left\{ \frac{i}{a}\left(b-\frac{1}{a-b}\right) \sinh(2h)
 +i\frac{Un^2}{4}\frac{a+b}{a-b}\right\}[{\cal L}_j,
   C_{\sigma,au}C_{\tau,au}]\label{lastterm}\\
&-&2\left\{ \frac{i}{a}\frac{a}{a-b} \sinh(2h)
 -i\frac{Un^2}{4}\frac{2ab}{a-b}\right\}I_{au}\left(U^{(-)}_{\sigma,j}
 (S^{(-)}_{\tau,j}+T^{(-)}_{\tau,j})+U^{(-)}_{\tau,j}
 (S^{(-)}_{\sigma,j}+T^{(-)}_{\sigma,j})\right)I_{au},\nonumber
\end{eqnarray}
where $a =\cos (\lambda)$, $b=\sin (\lambda)$.
In  deriving  equation (\ref{lastterm}), use has been made
of the following identities
\begin{eqnarray}
&&(a-b)[{\cal L}_j, \hat{C}_{\sigma,j}\hat{C}_{\tau,j}]
+(a+b)[{\cal L}_j,
   C_{\sigma,au}C_{\tau,au}]\nonumber\\
&=& -4ab I_{au}\left(U^{(-)}_{\sigma,j}
 (S^{(-)}_{\tau,j}+T^{(-)}_{\tau,j})+U^{(-)}_{\tau,j}
 (S^{(-)}_{\sigma,j}+T^{(-)}_{\sigma,j})\right)I_{au},\label{lastiden}\\
I_{au}Q_jI^{-1}_{au}&=&
   -\frac{i}{a}\sinh(2h)\left(U^{(-)}_{\sigma,j}{C}_{\tau,au}
+ U^{(-)}_{\tau,j}{C}_{\sigma,au}\right) -\frac{2i}{a}\sinh^2(h)\left(
U^{(+)}_{\sigma,j}+U^{(+)}_{\tau,j}\right),\nonumber
\end{eqnarray}
which are straightforward consequences of the identities
(\ref{idents}).
In  expression (\ref{lastterm}) let us choose the parameter $h$ by
\begin{equation}
\label{condition}
\sinh(2h)=\frac{Un^2}{4}2ab =\frac{Un^2}{4}\sin(2\lambda),
\end{equation}
so that the second  line in the r.h.s. of (\ref{lastterm}) vanish.
Then the Lax-pair can be written as
\begin{eqnarray}
 \frac{d{\cal L}_{j}}{dt}&=&\displaystyle
{\cal B}_{j+1}{\cal L}_{j}- {\cal L}_{j}{\cal B}_{j},
   \label{finallax}\\
{\cal B}_j & =&
{\cal M}_{j}+ Q_j
+i\frac{Un^2}{4}\left\{ 2b\left(b-\frac{1}{a-b}\right)
 +\frac{a+b}{a-b}\right\}C_{\sigma,au}C_{\tau,au}.\label{mmform}
\end{eqnarray}
Here, ${\cal M}_j, Q_j$ are defined by equations (\ref{calm})
and (\ref{q}), respectively. It is clear that the
condition (\ref{condition})
guarantees the parametrizability   for $SU(n)$ Hubbard model,
which was first introduced  in \cite{Maassarani2} through
the Yang-Baxter equation.
  As we know, the standard method to construct the open system is
to study the reflection equations \cite{Sklyanin}. The $R$ matrix
must enjoy unitarity and crossing symmetry.
For $SU(n)$ Hubbard model, the $R$ matrix in \cite{Maassarani2} does not
have the crossing symmetry and the invertibility of a related matrix is
lacking.
In the Lax pair formalism, however,  the detailed properties of the $R$
matrix do not come in.
 This is  another way to  study the integrability
of the open boundary systems. Therefore, the Lax-pair derived in the
present letter would be a useful tool for analyzing open systems.
We will consider its applications elsewhere.

\bigskip
\bigskip
\begin{center}
{\bf ACKNOWLEDGMENTS}
\end{center}
This work was supported by
Grant--in--Aid of Ministry of Education, Science and Culture of Japan.
R.Y. thanks the Japan Society for the Promotion of Science.

\end{document}